\documentclass{aa}
\usepackage{natbib,epsf,graphicx}
\bibpunct{(}{)}{;}{a}{}{,}
\let\tmpclearpage\clearpage
\let\clearpage\relax

\begin{document}

\title{Statistics of convective collapse events in the photosphere \\ and chromosphere observed with the Hinode SOT}
\author{C.E.~Fischer \inst{1} \and A.G.~de~Wijn \inst{2} \and R.~Centeno \inst{2} \and B.W.~Lites \inst{2} \and C.U.~Keller \inst{1}}
\institute{ Utrecht University, The Netherlands; C.E.Fischer@uu.nl; C.U.Keller@uu.nl  \and High Altitude Observatory, National Center for Atmospheric Research, Boulder USA; dwijn@ucar.edu; rce@ucar.edu; lites@ucar.edu}

\abstract
{}{ Convective collapse, a theoretically predicted process
  that intensifies existing weak magnetic fields in the solar
  atmosphere, was first directly observed in a single event by Nagata
  et al. (2008) using the high resolution Solar Optical Telescope
  (SOT) of the Hinode satellite. Using the same space telescope, we
   performed a statistical analysis
  of convective collapse events.} {Our data sets consist of high
  resolution time series of polarimetric spectral scans of two iron
  lines formed in the lower photosphere and filter images in
  Mg\,I\,b$_{2}$ and Ca\,II\,H, spectral lines that are formed in the 	
  high photosphere and the lower chromosphere, respectively. We were
  thus able to study the implication of convective collapse events on
  the high photospheric and the chromospheric layers. The physical parameters from the
  full Stokes profiles were obtained with the MERLIN Milne-Eddington
  inversion code.} { We found that in all cases, the event was
accompanied by a continuum bright point and nearly always by a
brightening in the Ca\,II\,H images. The magnesium dopplergram
exhibits a strong downflow in about three quarters of the events that
took place within the field of view of the magnesium dopplergram. For each of the
  49 events we determined the duration, maximum photospheric downflow,
  field strength increase and size. We found event durations of about 10 minutes, magnetic element radii of about 0.43~$\arcsec$ and 0.35~$\arcsec$, before and after the event, respectively, and field strengths of up to $1.65$~kG.}
{}

\titlerunning{Statistics of convective collapse events}
\authorrunning{C.E.Fischer et al. }

\maketitle

\section{Introduction}\label{sec:introduction}

\begin{figure}
\resizebox{\hsize}{!}{\includegraphics{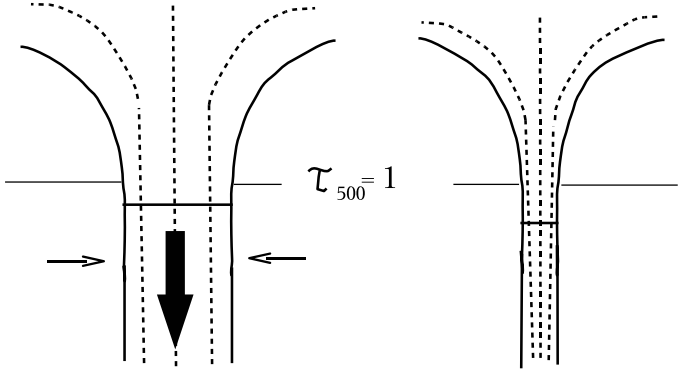}}
\caption{Convective collapse scenario according to
  \cite{1978ApJ...221..368P}.  On the left the magnetic flux tube is
  sitting in the intergranular lane and is strong enough to reduce
  convective heating, which will develop an adiabatic temperature
  gradient in its radiatively cooled atmosphere. The dotted lines
  represent the magnetic field and the horizontal line shows the
  $\tau$ = 1 level and the Wilson depression inside the flux tube. The
  convective collapse process assumes an initial mass downflow,
  represented by the big dark arrow pointing downwards. The
  temperature of the downwards traveling mass increases adiabatically,
  whilst the surrounding atmosphere shows a much steeper temperature
  gradient. This leads to an enhancement of the downdraft. The flux
  tube is then evacuated and forced to compress as the gas
  outside the flux tube is pressing horizontally
  against it, represented by the thin arrows. The limit of the compression is reached when
  the magnetic field pressure balances the outer gas pressure. The
  right side shows the compression of the flux tube, which leads to the
  intensification of the magnetic field strength.}
\label{carto}
\end{figure}

Time series of solar photospheric convection as seen in the continuum
at visible wavelengths and the corresponding magnetograms show a
highly dynamic quiet sun.  Movements, accumulation, and separation of
magnetic elements are clearly determined by the granular flows
\citep{1996ApJ...463..365B}. \cite{1973SoPh...32...41S} determined the
field strengths of the coalesced magnetic concentrations by using the
ratio between Stokes~$V$ profiles of two iron lines with roughly the
same excitation potential and line strength but with different
effective Land\'e factors.  This line-ratio technique revealed
that the spatial-resolution-independent field strength is around
1--2~kG.  This is significantly more than the equipartition field
strength of $\approx$ 500~G \citep{1999ApJ...522..518T}, which is
obtained by equating the known kinetic energy density of the
convective flows with the magnetic energy density. This led \cite{1978ApJ...221..368P} to a possible explanation for the increase in field strength and the observed kG fields. This scenario is illustrated
in cartoon form in Fig.~\ref{carto}.

Numerical simulations of the convective collapse process up to the
destruction of the stable kG flux tube have been carried out in 2\,D and 3\,D
\citep{1985A&A...143...39H,1999ApJ...522..518T,1998A&A...337..928G,2000ARep...44..701S}. The
work by \cite{1998A&A...337..928G} suggests that, in some cases, the
downward traveling mass can ``bounce of'' the lower layers in the
atmosphere and create an upward moving flow that could lead to the
destruction of the compressed flux tube. Another destruction mechanism
was suggested by \cite{2000ARep...44..701S} and involved the cancellation of opposite
polarity elements, destroying magnetic flux and permitting the
disintegration of the flux concentration.

After years of theoretical predictions and numerical simulations, the
first observational signs of the convective collapse scenario were
given by \cite{2001ApJ...560.1010B}, who studied asymmetric Stokes~$V$
profiles of two photospheric, near-infrared lines at fixed
spectrograph slit positions. They found redshifts in the Stokes~$V$
profiles, which correspond to the downflow phase during a convective
collapse process; however, shortly afterwards, they observed a
decrease in the magnetic field strength, after reaching only
600~G. They also observed at the same time a strong, 3-lobed
Stokes~$V$ profile and argued that this is due to the appearance of an
additional blueshifted Stokes~$V$ profile. This can be explained by a
sharp change from downflow to upflow along the line of sight and is in
agreement with an upward-traveling shock front. \cite{2001ApJ...560.1010B}
applied an inversion code based on the thin flux tube scenario
and confirmed that the profiles could be fitted very well
with a discontinuity in the line of sight velocity as the result of an aborted 
convective collapse event. The
first direct observation of a convective collapse event by
\cite{2008ApJ...677L.145N} shows a magnetic network element undergoing
a transition from a weak field strength of 400~G to a 2~kG flux tube.
The increase of magnetic field strength was preceded by a strong
downflow of 6 km\,s$^{-1}$. They were also able to link the
strong-field magnetic element to the formation of a continuum bright
point.  They concluded from these observations that they were
witnessing a convective collapse event.
We present new observations from the Hinode
Spectro-Polarimeter (SP) with the addition of chromospheric intensity
in Ca\,II\,H and Mg\,I\,b$_{2}$ dopplergrams revealing the dynamics of
the upper photosphere and perform a statistical analysis of 49
convective collapse events identified in several time series taken
with the Hinode SOT.

\begin{table*}

\caption{Statistics of convective collapse events (see section~\ref{sec:results}).}
\label{statstab}
\centering

	 	\begin{tabular}{l c c c c c c }
 Type of event&max v down [km\,s$^{-1}$]&start B [G]&peak B [G]&r 1 [arcsec]&r 2 [arcsec]&duration [min]\\
  \hline
 \hline
 11 without Mg downflow&3.80&819&1093&0.48&0.41&10.36\\
 28 with Mg downflow&3.92&886&1356&0.48&0.34&10.37\\
 10 Mg out of FOV&3.33&900&1262&0.34&0.31&9.09\\ 
 \hline	

		\end{tabular}

		\end{table*}

 \begin{figure}
\resizebox{\hsize}{!}{\includegraphics[width=12cm]{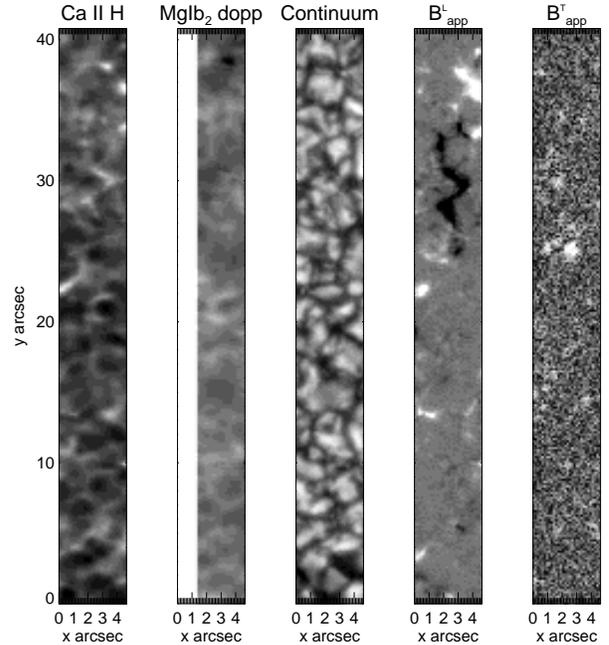}}
\caption{Snapshot of a time sequence from the reduced data.  From left
  to right, the coaligned images show the
  Ca\,II\,H intensity, the magnesium dopplergram, where black corresponds to
  downflows, the continuum intensity, the longitudinal and the transverse apparent
  magnetic flux (obtained by integrating over the circular and linear
  polarization profiles and multiplying by a known calibration
  constant). The left side of the Mg\,I\,b$_{2}$ dopplergram is cut
  off due to instrument misalignment and the restricted fields of
  view.}
\label{allpanels}
\end{figure}

\section{Data capture and processing}\label{sec:data}

We obtained 13 one-hour time sequences with the Hinode
Spectro-Polarimeter (SP) each covering an area of 41 by $4.8$ arcsec with a
cadence of 58 seconds and a pixel size of $0.16$ arcsec.  The Fe\,I
630.15~nm and Fe\,I 630.25~nm spectral lines were used with a spectral
resolution of $0.021$~\AA.  Simultaneous Narrowband Filter Imager
(NFI) images in the wings of Mg\,I\,b$_{2}$ and Ca\,II\,H filtergrams
with the Broadband Filter Imager (BFI) were recorded. The
Mg\,I\,b$_{2}$ filtergrams were taken in both the blue and red wings
at 114 m\AA\ with an exposure time of $0.41$~s and a pixel size of
$0.16$ arcsec. The Ca\,II\,H filtergrams were taken with an exposure
time of $0.15$~s and a pixel size of $0.11$ arcsec. For more
information on Hinode's Solar Optical Telescope see
\cite{2007SoPh..243....3K,2008SoPh..249..197S,2008SoPh..249..221S,2008SoPh..249..233I}.
We used the SolarSoft IDL routines sp\textunderscore prep and
fg\textunderscore prep to process the data to level 1. The
sp\textunderscore prep procedure corrects the SP data for dark
current, flat field, removes spectral and spatial drifts due to
thermal instability, and merges the data from the two SP beams. The
fg\textunderscore prep routine corrects the filtergram (FG) data for
dark current, flat field, and camera defects.

We then spatially aligned the FG and SP data. Due to the small field
of view in the narrowband filter and a fixed pointing offset between
FG and SP data, part of the SP field of view is outside the narrowband
filter field of view. The magnesium dopplergram therefore does not
cover the whole SP field of view. The FG data was resampled to the SP
timing using a nearest-neighbor algorithm, with the central slit of the SP
as a reference. 
The magnesium dopplergrams were obtained by dividing the difference
between the blue wing and the red wing images by the sum.
As the line-formation of the Mg\,I\,b$_{2}$ line is quite complicated and there is no simple
response height that can be assigned to the measured
doppler signal (M.Carlsson, private communication) we could not
obtain a calibration curve for the velocities in km\,s$^{-1}$ from observations or simulations.
However, the formation layer of the Mg\,I\,b$_{2}$ dopplergrams was
estimated by comparing calculated Mg\,I\,b$_{2}$ dopplergrams
with snapshots from MURaM simulations \citep[see code description]{2003AN....324..399V}, and the images revealed a strong similarity in the highest
photospheric layers (N.Vitas, private communication). 
In Fig.~\ref{allpanels} we show one scan from a time sequence.


We used the Milne-Eddington inversion code MERLIN, developed by the
Community Spectro-polarimetric Analysis Center \footnote{see http://www.csac.hao.ucar.edu}, to retrieve the
magnetic field parameters from the observed Stokes profiles
\citep{1987ApJ...322..473S,2007MmSAI..78..148L}.

\section{Results}\label{sec:results}

Convective collapse events were identified by eye in the observed data
and the inversion parameters. Each time series was checked for sudden
substantial increases in magnetic field strength preceded by a strong
photospheric downflow. We found that in all these cases, the event was
accompanied by a continuum bright point and nearly always by a
brightening in the Ca\,II\,H images. The magnesium dopplergram
exhibits a strong downflow in about three quarters of the events that
took place within the field of view of the magnesium dopplergram.
 
 Table~\ref{statstab} shows the statistics of the identified events in the 13 hours of
  repeated 4.8 x 41 arcsec${^2}$ raster scans. To obtain the
  properties of the magnetic elements we followed them by eye from one
  time step to the next. The maximum of the line-of-sight velocity and
  the magnetic field strength are averages over the three highest data
  points, where the starting magnetic field was taken at the point of the strongest downflow. The majority of the events show equipartition field strengths at about 130 seconds before that, but it is difficult to trace the magnetic element precisely at that time. The size was determined by tracing the magnetic element in
  the magnetograms by eye before (radius 1) and after (radius 2) its
  collapse. The duration of the event was determined by taking the
  time when the maximum downflow occurred as the start point until
  after the magnetic field had reached its maximum and declined to
  values comparable to equipartition field strength.

\begin{figure}
\resizebox{\hsize}{!}{\includegraphics{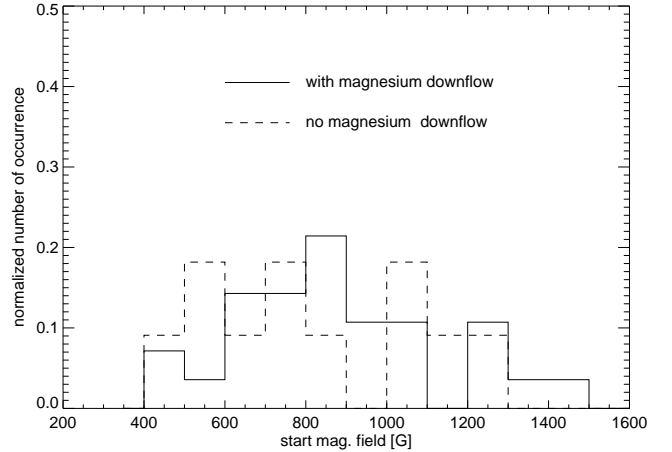}}
\caption{Histogram of the starting values of the magnetic field
  strength in the magnetic elements before the convective collapse
  showing the number of occurrence divided by the total number of
  events. The dashed lines correspond to events without a downflow in
  the magnesium dopplergram. The solid line represents the 28 events
  exhibiting downflows in the magnesium dopplergram.}
\label{Bbeghisto}
\end{figure}

\begin{figure}
\resizebox{\hsize}{!}{\includegraphics{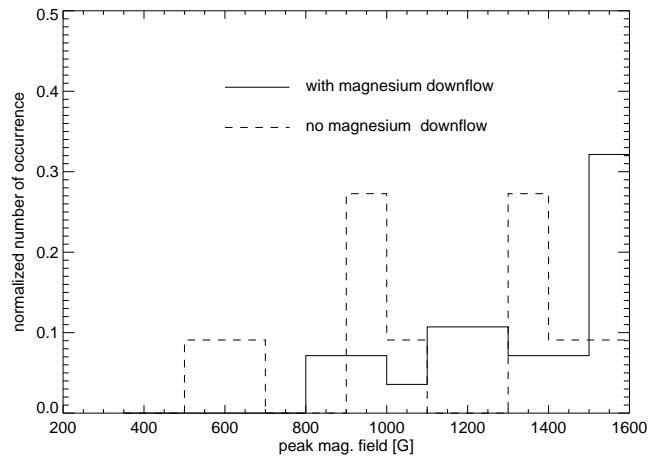}}
\caption{Histogram of the maximum field strength reached after the
  intensification in the same format as Fig.~\ref{Bbeghisto}.}
\label{Bendhisto}
\end{figure}

\begin{figure}
\resizebox{\hsize}{!}{\includegraphics{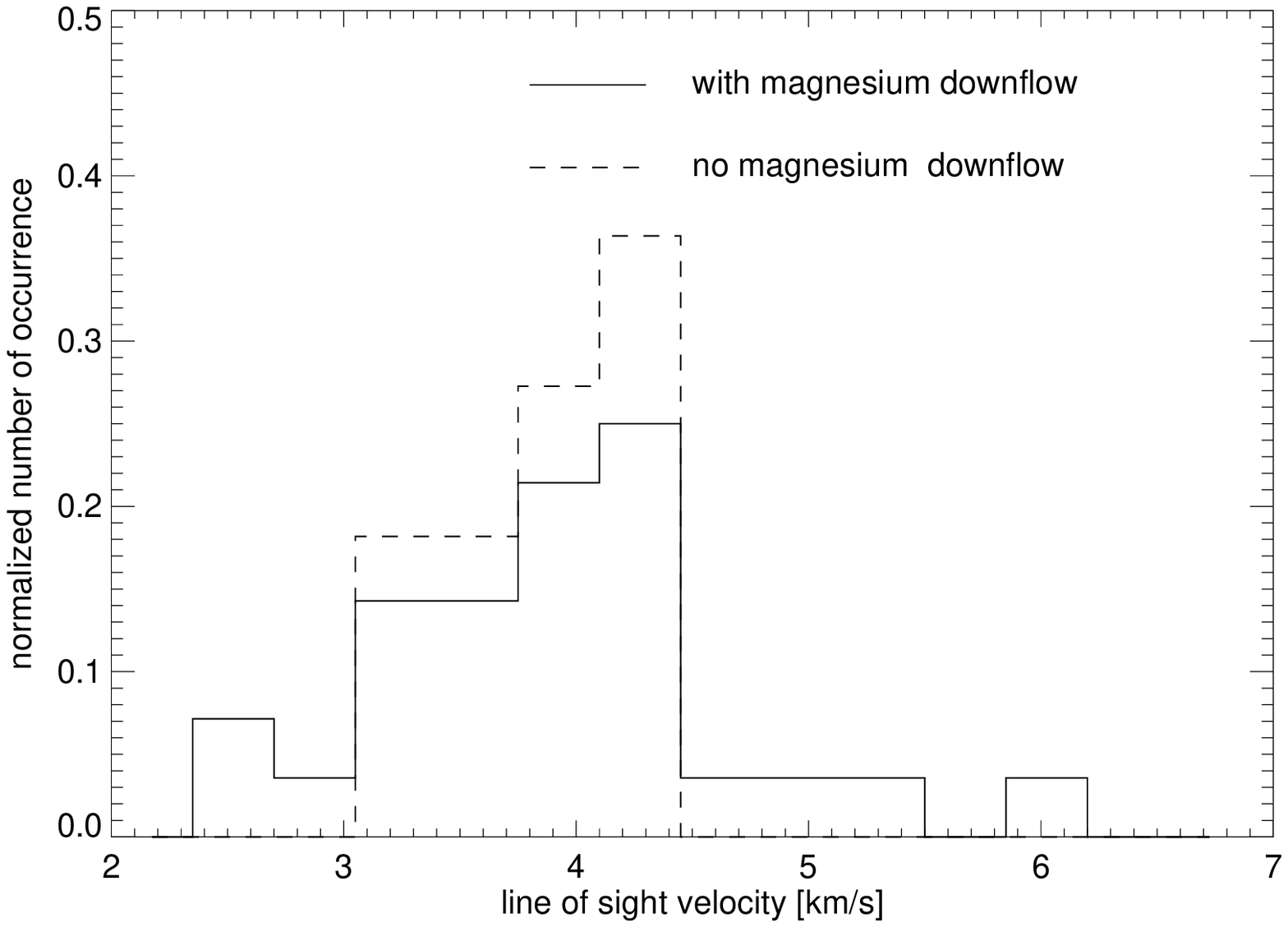}}
\caption{Histogram of the maximum photospheric line of sight velocity
  in the same format as Fig.~\ref{Bbeghisto}.}
\label{vlosfin}
\end{figure}

The values in Table~\ref{statstab} are averages, which might be misleading were the
distribution strongly asymmetric. The histograms of the line
of sight velocity and magnetic field strength are therefore shown in
Figures ~\ref{Bbeghisto},~\ref{Bendhisto} and ~\ref{vlosfin}.  These
histograms indicate that the distribution of events with downflows
seen in the high photosphere are similar to those obtained from the 11
events that do not exhibit downflows.

 \begin{figure*}
\sidecaption
\includegraphics[width=12cm]{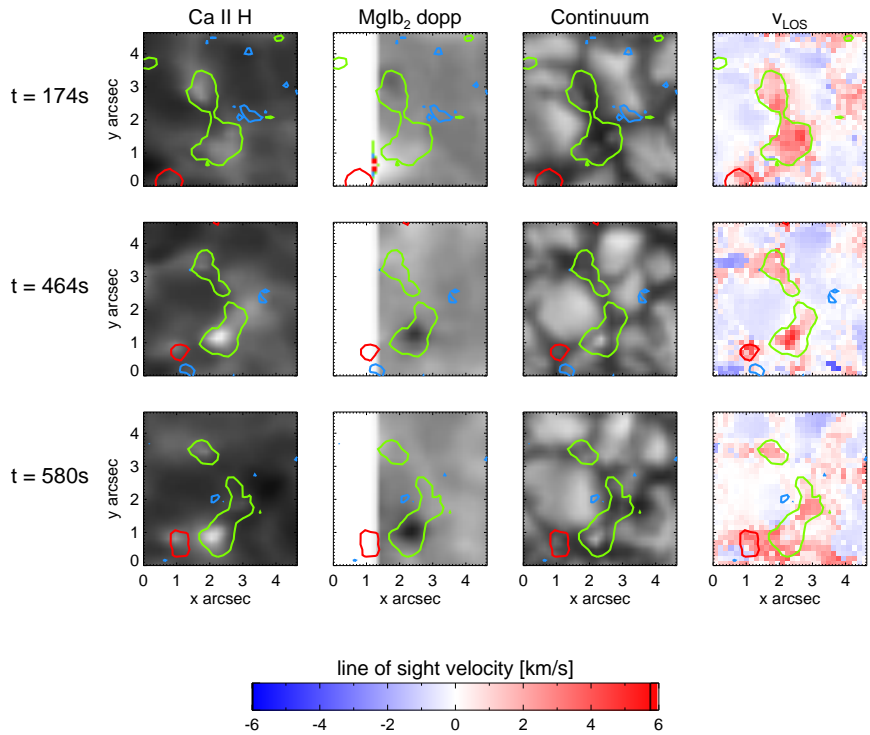}
\caption{A convective collapse event associated with downflows in the
  upper photosphere. The field of view is 4.8~$\arcsec$ by 4.8~
  $\arcsec$. The columns show the Ca\,II\,H intensity in a linear scale, the magnesium dopplergram in arbitrary units, the continuum
  intensity in a linear scale, and the line-of-sight velocity derived through inversion of the Fe I line profiles, with positive values corresponding to downflows. As discussed in section~\ref{sec:data} we do not have a calibration curve for the magnesium dopplergram. The units are therefore arbitrary and black corresponds to
  downflows in the magnesium dopplergram. Green contours indicate the location of
  line-of-sight flux directed toward the observer, while red contours indicate line-of-sight flux directed away from the observer.
  The contour levels are at $\pm\mathrm{30\,Mx/cm}^{2}$ in the
  apparent longitudinal flux density.  The blue contours indicate the
  transverse apparent flux density at $\mathrm{140\,Mx/cm}^{2}$.}
\label{evmag}
\end{figure*}

\begin{figure*}
\sidecaption
\includegraphics[width=12cm]{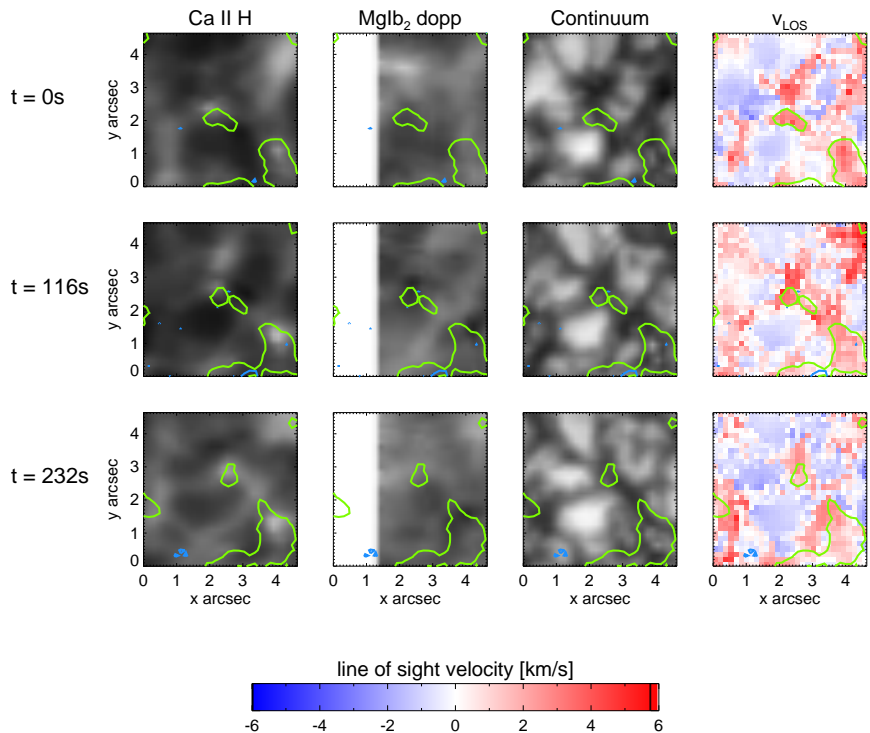}
\caption{A second example of a convective collapse event, in the same
  format as Fig.~\ref{evmag}. This example does not show a signature
  in the Mg\,I\,b$_{2}$ dopplergram.}
\label{evnomag}
\end{figure*} 

Figure~\ref{evmag} shows a typical sequence in the calcium intensity, magnesium dopplergram, where black
corresponds to downflows, the continuum intensity and the
photospheric line of sight velocity. Clearly, a bright point can be seen in
the continuum as well as in the Ca\,II\,H images forming just after a
strong photospheric downflow.  The magnesium dopplergram shows a
strong downflow, too. The flux tube is therefore evacuated not only in
the low photosphere but also in the high photosphere. As shown by the
statistical analysis, this is the case for most events.  An example of
an event without significant downflows in the Mg\,I\,b$_{2}$
dopplergram is shown in Fig.~\ref{evnomag}.

\begin{figure}
\resizebox{\hsize}{!}{\includegraphics{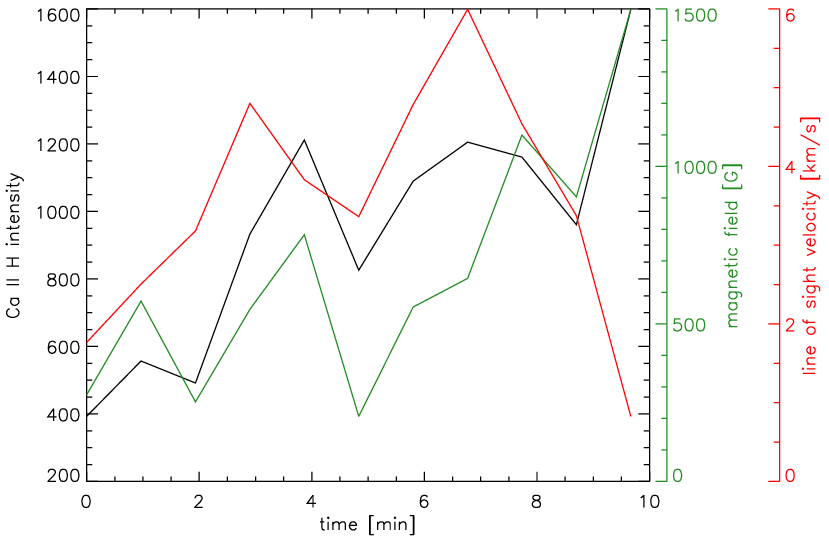}}
\caption{Temporal development of the Ca H II intensity, the line of
  sight velocity and the magnetic field strength of the magnetic
  element from Fig.~\ref{evmag}. Clearly, the maximum downflow takes
  place shortly before the increase in the Ca\,II\,H intensity and the
  magnetic field strength. The event duration in this case was longer than average, lasting about 15 minutes. We show here the first 10 minutes, in which the oscillatory behavior in each
  parameter is clearly evident.}
\label{timedep}
\end{figure}

The evolution of a case with magnesium downflow signals (c.f.\
Fig.~\ref{evmag}) is best seen in Fig.~\ref{timedep}, where we plot
the photospheric line-of-sight velocity, magnetic field strength in
the center of the magnetic element and the bright point intensity in
the Ca\,II\,H filter. There seems to be an oscillatory behavior in
each parameter. This is seen in most observed events and will be the subject of future research.

\section{Discussion}\label{sec:disc}

As seen in Table~\ref{statstab}, the duration of convective collapse
events is about 10 minutes. We can compare the duration of our events, i.e., the timespan 
from the moment of maximum downflow until the time that the field has again declined to 
equipartition strength, to the lifetime of magnetic elements, which has been determined
 by, e.g., \cite{2008ApJ...684.1469D}. They applied an automated feature-tracking algorithm to a six-hour
 sequence of magnetograms taken with the Hinode SOT, and found that magnetic elements on average have a lifetime of 10 minutes.
They note that magnetic elements intermittently drop below the detection limit of the algorithm.
It is tempting to speculate that the similarity between their element lifetime and our event 
duration results from the ability to track elements only while they are in a collapsed state.

 Another process showing similar observational signatures is the emergence of magnetic flux. This process has been well analyzed by, e.g., \cite{2002ApJ...569..474D,2007ApJ...666L.137C,2009arXiv0905.2691M}. \cite{2007ApJ...666L.137C} witness two opposite polarity patches emerging within a granular structure just after a horizontal field appears. We could be observing an emergent growing magnetic field instead of the enhancement of a field by the convective collapse process. However, the difference is that the events we examine take place in the intergranular lanes, and we can observe in most cases the accumulation of flux in the intergranular lane just before the process takes place. In the event shown in Fig.~\ref{evmag}, we can clearly see the development over minutes before the event takes place. The positive polarity flux (green contours) is carried by the granular flows and has been present in the intergranular lane during 20 minutes before the sudden increase of magnetic field. The opposite polarity patch near the observed event has been growing also through accumulation and has been also present in the intergranular lane for several minutes. It is therefore unlikely that we are observing a flux emergence event. The event in Fig.~\ref{evnomag} does not have an opposite polarity near it before or during the collapse.

\begin{figure}
\resizebox{\hsize}{!}{\includegraphics{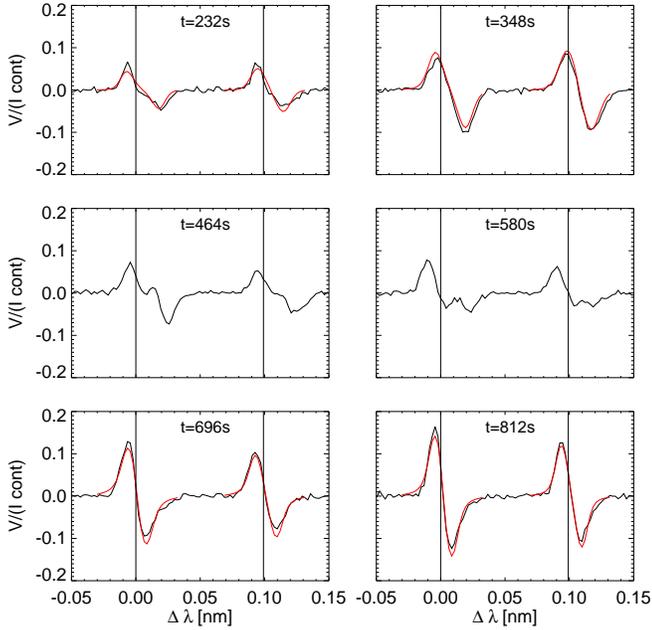}}
\caption{Temporal evolution of the normalized Stokes~$V$ profiles from
  the event in Fig.~\ref{evmag}. The snapshots are 116~s apart and the
  time runs from left to right and top to bottom. The vertical lines
  correspond to the position of the intensity minimum averaged over
  the whole map. We chose the pixel at the center of the magnetic
  element. The strongly distorted Stokes V profiles are seen after the maximal downflow and until the maximal field strength has been reached. This justifies the use of a  Milne Eddington code to obtain the line of sight velocity and the starting and peak magnetic field strength, as the Stokes profiles have a normal shape at these points. We show in red the fit obtained by the inversion code for the Stokes-V profiles at the beginning and end of the collapse.}
\label{vverl}
\end{figure}

It is still unclear what mechanism causes the destruction of the strong magnetic elements. According to
~\cite{2001ApJ...560.1010B}, an aborted convective collapse event can
cause upward traveling shocks, which manifest themselves as
asymmetries in the Stokes~$V$ profiles. The events observed here could fall
into this category and actually represent unfinished convective collapse
events, or could be weakened or destroyed by upward flows after the collapse, as seen in simulations
by \cite{1998A&A...337..928G}.  Taking a
closer look at the development of the Stokes~$V$ profiles
(Fig.~\ref{vverl}) of the event shown in Fig.~\ref{evmag}, we
observe a redshift caused by the downflowing material, but also strong
asymmetries and 3 lobes in Stokes~$V$ profiles in the center of the
magnetic element. They could be
signs for strong velocity gradients along the line of sight.

\begin{figure}
\resizebox{\hsize}{!}{\includegraphics{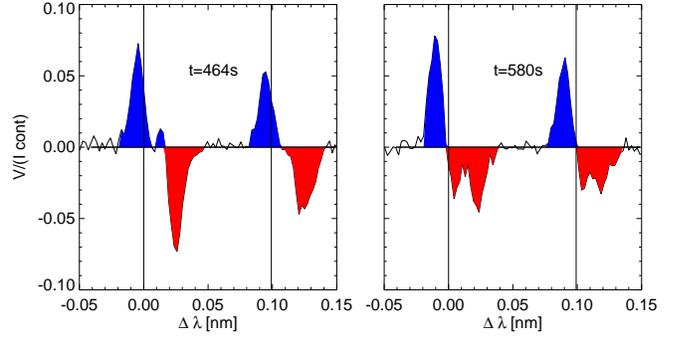}}
\caption{Two examples of strongly distorted Stokes V profiles observed
  within a kG magnetic element. The colors denote the blue or red
  lobe of the profile. We substract the blue area from the red area and
  divide by the sum to obtain the net circular polarization
  (NCP). All profiles show a NCP of less than 1~$\%$.}
\label{profV1}
\end{figure}

Figure~\ref{profV1} shows two examples of the observed asymmetric
Stokes~$V$ profiles, which can be either caused by several magnetic
concentrations with different velocities in the same resolution
element, or, as shown by \cite{1997ApJ...478L..45B}, magnetic elements with strong velocity and magnetic field
gradients along the line of sight.
In the first case the net circular polarization (NCP) should be zero
as the individual Stokes~$V$ profiles are antisymmetric. In the second
case there should be a substantial NCP (more than 2 percent). We
determined the NCP for several asymmetric Stokes~$V$ profiles, and find that all of
them are around or below 1 percent. We therefore cannot confirm these
events as being aborted convective collapse events or
upward-traveling shocks.

\begin{figure}
\resizebox{\hsize}{!}{\includegraphics{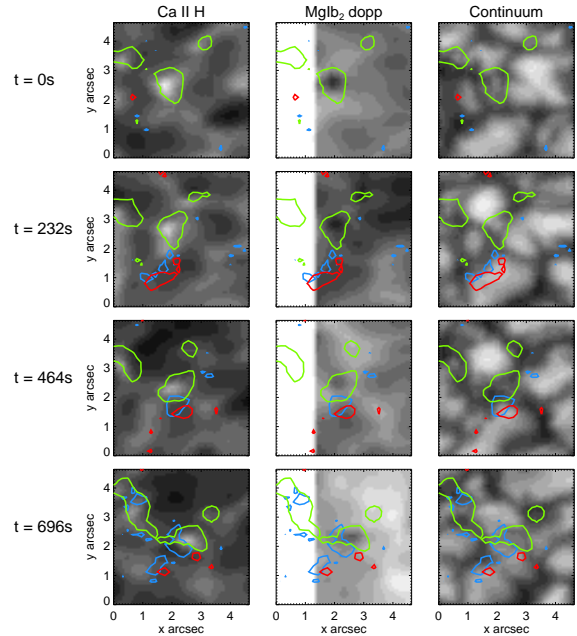}}
\caption{Flux cancellation event leading to the weakening of the
  collapsed magnetic element. The figure is in the same format as
  Fig.~\ref{evmag}, but without the photospheric line-of-sight velocity.}
\label{flcan}
\end{figure}

However, we did find examples consistent with another process that may
weaken the magnetic elements.  Figure~\ref{flcan} shows magnetic flux
of the opposite sign near to our investigated collapsed magnetic
element. The flux is carried by convective motions toward the magnetic
element. Then, horizontal flux is observed, and the inbound opposite
polarity flux disappears almost completely.  At the same time the
field strength decreases sharply within the collapsed magnetic
element. This flux cancellation scenario occurs several times in our
datasets and might be the leading cause for the destruction of the
newly formed kG fields. This is in agreement with simulations by
\cite{2000ARep...44..701S}, who state cancellation as the most common
process of flux dispersal. As mentioned previously, we have to consider that we might actually be observing an emerging flux event. We do observe in the event of Fig.~\ref{flcan} opposite polarity flux emerging and growing in the granular structure in the vicinity of the observed magnetic element, but at that point the convective collapse event has already taken place and a bright point is already visible in the continuum.

The Stokes profiles were fitted using a variable filling factor. As we are in the weak field regime, the product 
of the filling factor and the magnetic field strength is very well determined by the inversion, 
but the filling factor and the magnetic field strength on their own are significantly less well determined. 
We therefore also computed the magnetic flux density per pixel 
inside the magnetic element, i.e. filling factor multiplied by the magnetic field, and averaged 
these values. Comparing the magnetic flux density before and after the collapse we found an increase 
in the vast majority of events, which is in agreement with the convective collapse scenario. 
The exceptions may be a consequence of the difficulty in determining the correct size of the magnetic elements by eye. 

In theory, the total flux of the magnetic elements (on the order of
$10^{18}$ Mx) should not be altered by the convective collapse as the
increase in magnetic field strength is balanced by the decrease in
area covered by the magnetic field. We do, however, find a systematic
decrease in magnetic flux, which may indicate a problem with
determining the flux, the field strength, the filling factor, a real
decrease in flux, or a combination of all of the above.

\section{Conclusion}\label{sec:conc}

We found 49 convective collapse events in the quiet solar photosphere,
all showing a brightening in the continuum intensity and reaching peak
magnetic field strengths of up to $1.65$~kG, and with almost all events
showing a brightening in the Ca\,II\,H intensity. The evacuation of
the flux tube is demonstrated by downflows in the lower photosphere, which are
also observed in the upper photospheric layers. Our observations provide very
clear evidence of the ubiquitous occurrence of the convective collapse
process. Only about a quarter of these events do not show a
significant downflow in the high photosphere. We also found that the
duration of the events from the downflow until the peaking and
subsequent decrease of magnetic field strength is about 10 minutes. 
In addition we confirm flux cancellation as a promising candidate for
the weakening of the magnetic field strength in these strong field
elements as predicted by simulations described by
\cite{2000ARep...44..701S}.

\begin{acknowledgements}
  We thank Nikola Vitas for estimating the formation height range of
  the Mg\,I\,b$_{2}$ dopplergrams and Rob Markel for providing the
  MERLIN inversions through the HAO inversion client.  

This research
  project has been supported by a Marie Curie Early Stage Research
  Training Fellowship of the European Community's Sixth Framework
  Programme under contract number MEST-CT-2005-020395: The USO-SP
  International School for Solar Physics.  

The National Center for
  Atmospheric Research is sponsored by the National Science
  Foundation. 

Alfred de\,Wijn and Catherine Fischer acknowledge travel support through the NCAR Early Career Scientist Assembly visitor fund.

Hinode is a Japanese mission developed and launched by
ISAS/JAXA, with NAOJ as domestic partner and NASA and STFC
(UK) as international partners. It is operated by these agencies in
co-operation with ESA and NSC (Norway).
\end{acknowledgements}

\let\clearpage\tmpclearpage
\bibliography{cfischerarxiv}
\bibliographystyle{aa}

\end{document}